# CrySPAI: A new Crystal Structure Prediction Software Based on Artificial Intelligence

**Zongguo Wang [1,2,\*], Ziyi Chen [1,2] , Yang Yuan, and Yangang Wang [1,2]**

[1]  Computer Network Information Center, Chinese Academy of Sciences, Beijing, 100083, China

[2]  University of Chinese Academy of Sciences, Beijing, 100049, China

\*  Correspondence: wangzg@cnic.cn

**Abstract:** Crystal structure predictions based on the combination of first-principles calculations and machine learning have achieved significant success in materials science. However, most of these approaches are limited to predicting specific systems, which hinders their application to unknown or unexplored domains. In this paper, we present CrySPAI, a crystal structure prediction package developed using artificial intelligence (AI) to predict energetically stable crystal structures of inorganic materials given their chemical compositions. The software consists of three key modules, an evolutionary optimization algorithm (EOA) that searches for all possible crystal structure configurations, density functional theory (DFT) that provides the accurate energy values for these structures, and a deep neural network (DNN) that learns the relationship between crystal structures and their corresponding energies. To optimize the process across these modules, a distributed framework is implemented to parallelize tasks, and an automated workflow has been integrated into CrySPAI for seamless execution. This paper reports the development and implementation of AI AI-based CrySPAI Crystal Prediction Software tool and its unique features.

## 1. Introduction

Crystal structure plays a fundamental role in understanding the physical and chemical properties of solid materials. One of the key challenges in materials research is how to efficiently and accurately determine crystal structures. Currently, two primary approaches are used to obtain target structures for inorganic materials. The first approach involves searching for similar structures in established crystal structure databases, such as the Inorganic Crystal Structure Database (ICSD)[1], the Pauling File[2], and others. The second approach involves predicting crystal structures by substituting elements in structure prototypes using high-throughput techniques[3-5]. While these methods can sometimes yield accurate structures quickly, they are less effective when dealing with new or unknown structure types. Moreover, the extremely high computational cost associated with the substitution method limits its practical applicability. Consequently, the ability to rapidly predict crystal structures based solely on chemical composition remains a pressing issue in theoretical materials research.

In recent years, several computational methods have been proposed and widely used for crystal structure prediction. These include the simulated annealing algorithm (SA)[6], genetic algorithm (GA)[7-11], particle swarm optimization (PSO)[12-13], and ab initio random structure searching (AIRSS)[14]. These methods have achieved notable success in predicting the structures of element, binary, ternary systems. Building on these approaches, several software packages for structure prediction have also been developed. Notable examples include USPEX (Universal Structure Predictor: Evolutionary Xtallography), which uses GA method[7-8], CALYPSO (Crystal structure AnaLYsis by Particle Swarm Optimization), which is based on PSO method[13], an adaptive-GA method that combines classical potential and DFT calculations[11], and XTALOPT, which implements an evolutionary algorithm with hybrid operators[15]. In all of these packages and workflows, structural energies are typically calculated using density functional theory (DFT). While DFT provides accurate results, its high computational cost limits the size and complexity of the structure cells that can be modeled. In addition, the accuracy of DFT is influenced by the choice of exchange-correlation functional and basis set, which can introduce uncertainties in energy rankings and affect the prediction of the most stable structures. In contrast, artificial intelligence technologies offer the advantage of lower computational costs and shorter development cycles, presenting a promising alternative.

With the increasing performance of artificial intelligence across various fields, machine learning, coupled with powerful DFT data, has become increasingly common in materials design and discovery[16-23]. In particular, interatomic potentials trained by machine learning have been employed to predict thermodynamical and other properties of bulk materials, achieving DFT-level accuracy for energy calculations. The typical models include structure-property relationship model and force field model. Structure-property relationship mainly includes Crystal Graph Convolutional Neural Network (CGCNN)[24], MatErials Graph Network (MEGNet)[25], and Tripartite interaction representation algorithm-enhanced Crystal Graph Neural Networks (TiraCGCNN)[26]. The widely used force field model includes M3GNet[27], CHEGNET[28] and GPTFF[29]. To facilitate their use, several software packages have been developed to construct accurate atomic interaction potentials. For example, the open-source Atomic Energy Network (aenet)[30] provides a deep learning-based representation of potential energy. DeepKit also provides the representation of potential energy and force fields, enabling molecular dynamics simulations[31-32]. Additionally, workflows have been developed to generate new crystal structures. Notable examples include MAGUS, which combines DFT calculations with machine learning methods[33], and a framework that integrates a graph network model with an optimization algorithm[34]. A more ambitious effort by Google DeepMind utilized AI to predict 380,000 new materials with the GNoME model[35]. While these existing tools and programs have demonstrated success in predicting crystal structures within predefined chemical or structural families, they face significant challenges when applied to unknown or unexplored materials. Therefore, the development of a new program is urgently needed to overcome these limitations by introducing a generalized, robust, and efficient approach.

In this paper, building on our previous work with the adaptive genetic algorithm (AGA)[36-37], we propose a crystal structure prediction software based on artificial intelligence (CrySPAI), which supports automatic structure optimization by combining AI technology with DFT data. The software consists of three modules: an evolutionary optimization algorithm (EOA) for searching crystal structure configurations, DFT calculations for determining energy values, and a deep neural network (DNN) for fitting the relationship between structures and their energies. Additionally, a structure-energy function is employed to filter out unreasonable structures in the EOA module. CrySPAI offers four distinct advantages, making it a powerful tool for researchers in the field of materials science: (1) Broad applicability. It can be applied across a wide range of inorganic materials. (2) Seamless integration and automation. The software integrates and automates all stages of the prediction process. (3) Enhanced predictive accuracy and efficiency. By combining AI and DFT, CrySPAI delivers improved accuracy and efficiency. (4) Exploration of unknown domains. It provides robust capabilities to explore materials in previously uncharted domains.

## 2. Results

### 2.1. Framework of CrySPAI

Three modules have three main modules, EOA, DFT and DNN. The schematic workflow of CrySPAI is shown in Figure 1. The input consists of the chemical composition provided by the user, and the output is the recommended crystal structure that CrySPAI determined to be energetically stable. The iterations in the middle shown in Figure 1 is used to train a convergence model, and it is the core of CrySPAI. The accuracy of model, as well as its applications, are key factors in determining the overall performance of the software. The model is iteratively trained to accurately predict the energy of different structures. The EOA module and DFT module in the iterative loop provide samples for the data set of the model training, while the EOA module operates outside the iterations to search for reasonable structures based on the trained model. Two databases are employed: one stores the results from DFT calculations, and the other contains the converged models.

Model training is a time-consuming and continuously evolving process, particularly when exploring new materials from scratch. However, once the model has been successfully trained, it facilitates efficient structure searching by comparing real-time structural energy information. This process, shown as the right process in Figure 1, enables CrySPAI to generate numerous satisfactory output structures, optimizing the search for the most stable configurations.

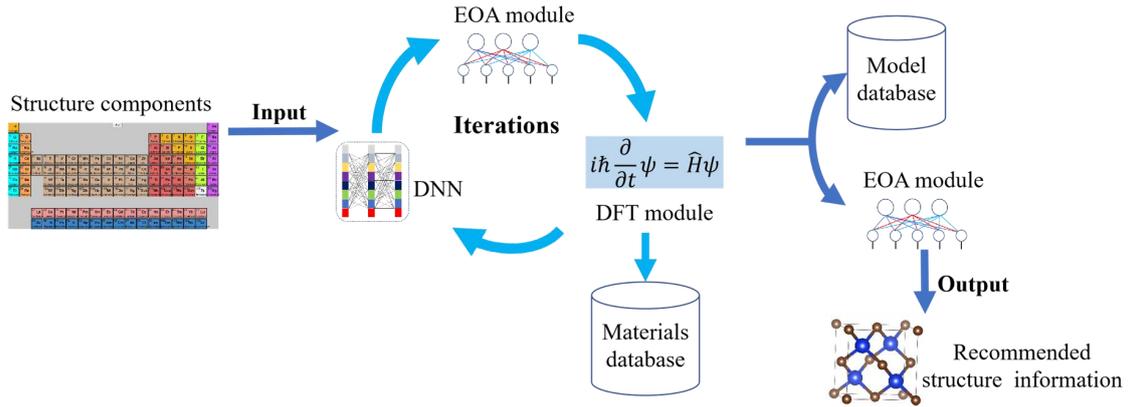

**Figure 1.** Schematic workflow of CrySPAI, showing the EOA, DFT, and DNN modules, along with databases for storing DFT results and model parameters.

*2.2. Applications on Crystal Structure Prediction of CrySPAI*

The simplest way to predict crystal structures using CrySPAI is by providing the component information of the structure, such as the element names and their atomic numbers. If you wish to modify the default settings of CrySPAI, you can do so by editing the parameter file. The default parameter file, args.py, includes structure information, model training parameters, and computational resource settings.

2.2.1 Parameters and Files Preparation

For the structure information, five main parameters need to be specified. The parameter names, default values, and descriptions are listed in Table 1.

**Table 1.** Parameters to construct crystal structure for CrySPAI.

| name | description | default value | type |
|---|---|---|---|
| entry 1atom | element | no default value | string list |
| atomn | number of each atom | no default value | int list |
| avolume | atomic volume of structure | calculated by equation 2 | float list |
| aenergy | atomic energy | atomn*0.0 | float list |

| crystalsys | crystal system of target structure | all | string |
| --- | --- | --- | --- |

Among these parameters, *atomType* and *atomn* are required, while the others are optional. The *aenergy* parameter is used to determine whether a new structure is stable. If the energy of the new structure exceeds the sum of the *aenergy* values for all atoms, CrySPAI will consider the structure unstable and discard it; otherwise, it will be retained. The *crystalsys* parameter specifies the crystal system of the target structure. However, throughout the training process, seven crystal systems are used to develop a universal model, as mentioned earlier.

In addition to the structure parameters, some input calculation files must be prepared in advance. The specific files required depend on the calculation software accessed by CrySPAI's DFT module, and users are free to customize the settings of these files. For the Vienna Ab initio Simulation Package (VASP), the required files include INCAR and POTCAR. If these input files are not provided, CrySPAI will automatically generate an INCAR file with default parameter settings. In the INCAR file, the "Accurate" precision mode is selected to avoid aliasing or wrapping errors. A conjugate-gradient algorithm is used to relax the ions to their instantaneous ground state, with a maximum of 100 ionic steps and an energy convergence criterion of $10^{-6}$ eV/atom. A plane-wave basis set is employed, with the kinetic-energy cutoff determined by the precision mode or manually specified by the user. The smallest allowed spacing between k-points is 0.1 Å$^{-1}$. In this work, the generalized gradient approximation (GGA)[38], specifically the Perdew-Burke-Ernzerhof (PBE) form, is used to calculate the exchange-correlation energy. While PBE is widely used for inorganic materials, we acknowledge its limitations and recommend users validate results for specific systems using alternative functionals when necessary. CrySPAI allows users to specify any functional supported by the chosen calculation software, such as LDA, meta-GGA, hybrid functionals, or even dispersion corrections like DFT-D2 or DFT-D3. Users performing DFT calculations with VASP must hold a valid VASP license. CrySPAI also supports other first-principles calculation software, such as Abinit, providing users with flexibility in their computational workflows.

Parameters related to model convergence and the training method are specified in the arg.py file, and users can review and modify them as needed. Once all preparations are complete, users can configure computational resources and submit tasks to supercomputers. After all iterations are finished, the results, including DFT calculations and trained models, will be stored in the database, and the recommended structure information will be output in a file format.

2.2.2 Crystal Structure Prediction

To demonstrate the effectiveness of the software in structure prediction, we selected several widely studied representative materials, including the electronic device material Si, the photocatalytic material $TiO_2$, the perovskite material $CaTiO_3$, and the metallic material Mg. These materials serve as foundational structures for a wide range of research areas, and their modifications have led to numerous research hotspots. The accuracy of structure prediction for these materials highlights the application potential of our software. Moreover, these representative materials also showcase CrySPAI's capability to predict structures for elemental crystals, binary compounds, and ternary compounds.

We specified the atomic numbers and volumes for Si, $TiO_2$, Mg, and $CaTiO_3$ to generate their crystal structures. The composition and volume information for these materials are provided in Table 2. All models are trained from scratch based on DFT calculation results, with structures generated at a fixed volume. These structures are derived from the seven crystal systems, and the initial model is trained using DFT results from randomly generated structures. The model is then iteratively updated until it converges globally. The EOA module uses the converged model to search for target structures.

**Table 2.** Structure information of Si, $TiO_2$, Mg, and $CaTiO_3$. Comparison of experimental and predicted space groups, c/aBias, and siteBias.

| structure | atomType | atomNumber | volume of Cell($Å^3$) | TargSpg /PredSpg | c/aBias | siteBias |
|-----------|----------|------------|-----------------------|------------------|---------|----------|
| Si | Si | 8 | 160.1 | 227/227 | 0 | 0.175 |
| $TiO_2$ | O,Ti | 8,4 | 136.28 | 227/227 | -0.259 | 0.258 |
| Mg | Mg | 2 | 45.41 | 194/194 | 0.048 | 0.440 |
| $CaTiO_3$ | Ca,O,Ti | 1,3,1 | 59.17 | 194/194 | 0 | 0.419 |

To describe the local atomic environment, the feature vector is constructed using a Chebyshev expansion[39], ensuring that the computational complexity of the machine learning model remains manageable even as the number of chemical species increases. Default DFT calculation parameters are employed. The model convergence tolerance is set to 8 meV/atom, and the 'patience' for epochs with no loss improvement is set to 40, based on repeated testing. Additionally, the swarm algorithm is integrated to enhance the prediction accuracy of the neural network during the training process.

The predicted structure information, including the space group, lattice parameters, and the biases in atomic positions between the predicted and experimental structures, are summarized in Table 2. CrySPAI generates multiple candidate structures for each material, including known polymorphic forms. In Table 2, we present one of the 16 lowest-energy structures for each material to demonstrate that CrySPAI successfully identifies experimentally observed stable phases. Users can access the complete ranked list of predicted structures to explore other stable or metastable configurations. The c/aBias represents the difference in shape between the predicted and experimentally stable crystal structures. The site bias is calculated as the root mean square error (RMSE) of atomic site positions. From Table 2, it can be observed that crystal structures with orthogonal angles exhibit higher accuracy, particularly in terms of atomic coordinate positions. In contrast, the prediction errors for structures with inclined geometries are larger due to the combined deviation in both lattice constants and site positions. Additionally, the space group can be accurately predicted when the volume is fixed, and the smaller site bias values are observed in these cases. These findings demonstrate that CrySPAI exhibits strong predictive power for commonly used materials, particularly in identifying stable crystal phases with high precision.

## 3. Discussion

For an unknown structure, CrySPAI first calculates its volume and then searches for the structural information, including shape and atomic positions. In all these processes, the accuracy of the model is crucial. The network model quickly predicts the energies of structures and recommends the optimal structural candidates for further generation. To enhance the robustness and efficiency of CrySPAI, an adaptive volume adjustment algorithm and a hybrid swarm intelligence algorithm are introduced. The adaptive volume adjustment algorithm is employed to determine an appropriate unit cell, while the hybrid swarm intelligence algorithm —combining the strengths of genetic algorithms, particle swarm optimization, and Bayesian optimization—is used to optimize the neural network, improving model stability and training efficiency. Further details of this method can be found in our previous work[40].

*3.1. Volume Prediction Performance*

To verify whether CrySPAI can accurately predict a suitable volume for the structure, we first calculated the volumes of five typical metal crystal structures to test the adaptive volume adjustment algorithm we developed. Table 3 presents the predicted and experimental results for these five metal elemental crystals, along with their respective volumes. As shown, the volumes of the structures recommended by CrySPAI are generally consistent with the experimental standard values, with a volume error of less than 3 $Å^3$/cell. This demonstrates that the proposed volume adjustment strategy is effective for searching unknown structures. Additionally, the space groups were predicted with good accuracy, particularly for $Ni_4$, which shows a higher level of precision.

**Table 3.** The predicted volume for different structures.

| Structure | TargVol/PredVol($Å^3$) | TargSpg/PredSpg | tolerance($Å$) |
|---|---|---|---|
| $Cu_4$ | 50.007/47.238 | 225/225 | 0.2 |
| $Ni_4$ | 42.842/43.774 | 225/225 | 0.01 |
| $Mg_2$ | 45.405/46.454 | 194/194 | 0.1 |
| $Zn_2$ | 30.319/29.792 | 194/194 | 0.2 |
| $Zr_2$ | 46.299/46.570 | 194/194 | 0.1 |

*3.2. Swarm Intelligence Algorithm Performance*

To evaluate the performance of the swarm intelligence algorithm in CrySPAI, we tested model convergence using a small dataset and conducted structure searches based on trained models for Li, Ca, and Mn. The results were compared with those obtained using the traditional back-propagation algorithm, as shown in Table 4. In this test, the total dataset consisted of approximately 1000 structures, the loss was kept below 0.08 eV/atom, and the iteration count was set to 1500 for each loop of the EOA module used for structure searching.

**Table 4.** Comparison between swarm intelligence algorithm and and traditional back-propagation algorithm in the process of model training and structure searching.

| Structure | OptimizedMethod | loss | dataset size | loop number |
|---|---|---|---|---|
| Li | back-propagation algorithm | 0.074 | 668 | Null |
| | swarm intelligence algorithm | 0.009 | 673 | 1 |
| Ca | back-propagation algorithm | 0.075 | 1279 | Null |
| | swarm intelligence algorithm | 0.053 | 1216 | 1 |
| Mn | back-propagation algorithm | 0.042 | 1135 | Null |

| | swarm intelligence algorithm | 0.05 | 1335 | 2 |
|---|---|---|---|---|

As shown in Table 4, the proposed swarm intelligence algorithm demonstrates excellent performance in structure searching. It effectively utilizes small datasets to achieve rapid convergence and exhibits strong generalization ability. In comparison, while the traditional back-propagation algorithm also performs well with small datasets, it is more prone to getting stuck in local optima or overfitting, which can hinder the EOA module from finding the optimal structure during the search process. Therefore, the swarm intelligence algorithm in CrySPAI offers faster convergence and higher accuracy, as the recommended individuals provide a stronger initial advantage. Additionally, energy prediction models for $TiO_2$ and $CaTiO_3$ were trained in earlier tests, showing that the neural network optimized by the proposed hybrid swarm intelligence algorithm improves both efficiency and accuracy, particularly for systems with a larger variety of atoms[40].

*3.3. Capability Performance of CrySPAI in Crystal Structure Search*

To evaluate the performance of CrySPAI, we compare it with two well-known structure search software, CALYPSO and GSGO, using available data from previous studies[12]. CALYPSO employs a particle swarm optimization algorithm, while GSGO utilizes a genetic algorithm for structure search. Both software packages use first-principles calculations for optimization. It is important to note that each software searches different structural systems independently, so we treat the search in each system as one generation. When comparing the same optimized population size $N_{pop}$, CrySPAI reproduces experimentally reported structures in fewer generations than the other methods shown in Table 5. This improvement is primarily due to the efficiency of the DNN model, which reduces computational cost, and the more accurate parent structures provided by the model, allowing CrySPAI to more effectively search for the global optimum.

**Table 5.** Comparison between CrySPAI and other methods for several structure systems with equal population sizes($N_{pop}$).

| Structure | Algorithm | Prototype tructures | Generations | $N_{pop}$ |
|---|---|---|---|---|
| | CALYPSO | Diamond | 8//5 | 16 |
| Si | GSGO | Diamond | 15 | 16 |
| | CrySPAI | Diamond | 2 | 16 |
|     | CALYPSO  | Zinc blende | 8//5   | 12 |
| SiC | GSGO     | Zinc blende | 5      | 12 |
|     | CrySPAI  | Zinc blende | 1      | 12 |
|     | CALYPSO  | Zinc blende | 16//5  | 12 |
| GaAs| GSGO     | Zinc blende | 19     | 12 |
|     | CrySPAI  | Zinc blende | 2      | 12 |

## 4. Materials and Methods

### 4.1. Implementations of CrySPAI

#### 4.1.1. EOA Module

The GA, inspired by Darwinian evolution, has been widely used for crystal structure optimization. In CrySPAI, the EOA module employs GA to generate structures through inheritance, mutation, selection, and crossover operations. To enhance the comprehensiveness and accuracy of the structure search, the EOA module runs seven parallel procedures, each corresponding to a different crystal system. Each procedure consists of several iterations using the same GA operations. The workflow of the EOA module for the cubic system is shown in Figure 2.

The input information typically includes the element names, number of atoms, and optional volume data. When the EOA module receives the input for the target materials, it activates seven parallel processes, each corresponding to a different crystal system. For example, Figure 2 illustrates the structure search process for the cubic crystal system. Initially, some candidate crystal structures are generated based on the input data with the help of the ICSD database. These structures are then subjected to energy calculations in the "local optimization process." During this process, the energies of the crystal structures are predicted using a trained model. If the model is unavailable, energy calculations based on empirical formulas or single-point DFT calculations are used instead.

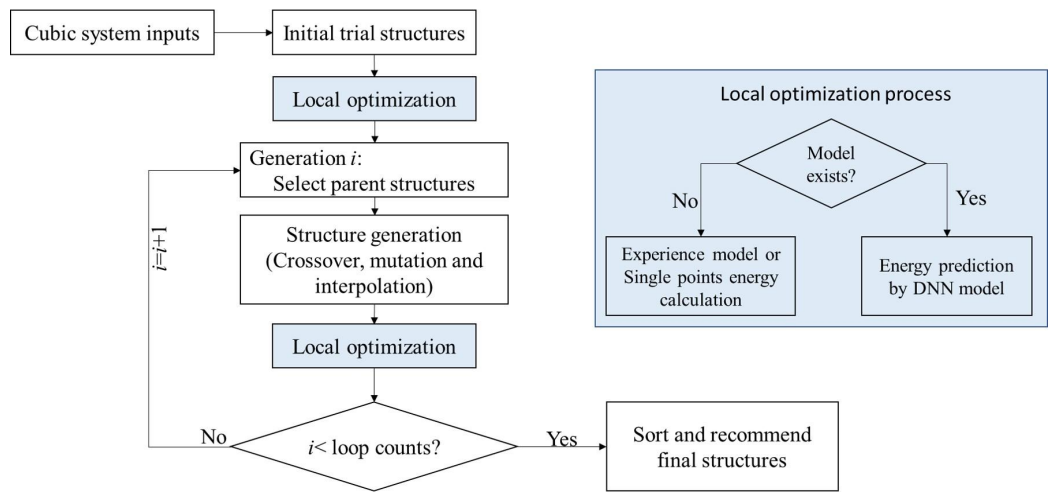

**Figure 2.** Structure search flowchart of the EOA module for the cubic crystal system, with "Local Optimization" for selecting stable structures from GA outputs.

After the first "local optimization process," both the structures and their corresponding energies form the trial structure set. The GA then begins its operation. The default number of trial structures in each generation $N_{trial}$ is 64. One-third of these, the structures with the lowest energy values, are selected as parent structures for crossover and mutation to produce the next generation. In each generation, 16 optimal structures with the lowest energy values are selected and interpolated into the trial structure set, replacing the 16 structures with the highest energy values from the previous generation. Once the GA generation loop ends, $N_p$ structures are recommended and output to the user as the final results, with the default value of $N_p$ being 16 for each crystal system.

If the EOA module is part of the iteration loop of the DNN module, to ensure the generalization ability of the model, 112 recommended structures (corresponding to seven crystal systems) from each GA generation loop are always transferred to the DFT module, regardless of whether the crystal system of the target structure is specified. Once the model is trained, the structure search for the specified crystal system will be conducted separately in the EOA module, and the default number of recommended structures is 16.

### 4.1.2 DFT Module

The DFT module is used to obtain accurate energy values for crystal structures. These results, along with their corresponding structures, form the training set for the model. As the choice of exchange-correlation functional and basis set can influence the energy calculations and ranking results, a brief note has been included to draw the user's attention to this critical aspect. In addition to its role within the iteration loop of the DNN module (as shown in Figure 1), the DFT module can also be used to calculate the energies of structures recommended by the EOA module upon user request. The DFT module in CrySPAI interfaces with external DFT calculation software, with VASP being the default computational tool. Input files such as POTCAR (containing pseudopotential information) and INCAR (with calculation parameters) must be prepared by the user, while the POSCAR file is automatically generated and transferred from the EOA module. The KPOINTS file is optional. Given the time-consuming nature of DFT calculations, parallel computing and high-throughput computational techniques are employed to accelerate the process. All relevant information, including composition, lattice parameters, atomic positions, and their energies, is extracted and stored in a MongoDB database in a standardized format for each DFT calculation.

The goal of the DFT module is to construct a comprehensive training dataset, which is continuously updated and expanded until the DNN iterations are complete. All results from DFT calculations are stored in the Materials database; however, not all of these calculations are added to the training set. CrySPAI considers two cases in which structure-energy pairs are excluded from the training set. The first case involves structures that are similar to those already present in the training dataset. The second case involves "useless" structures, where the energies from DFT calculations closely match the predicted values from the existing potential model, indicating little new information is provided.

### 4.1.3 DNN Module

Once the number of DFT calculations exceeds a specific threshold, the DNN module is activated. An atomic energy DNN model is then generated, and the model undergoes continuous training and updates in the model iteration loop, as shown in Figure 1, until convergence is achieved. The default neural network used in CrySPAI consists of four layers: one input layer, two hidden layers, and one output layer. The architecture of the neural network is illustrated in Figure 3.

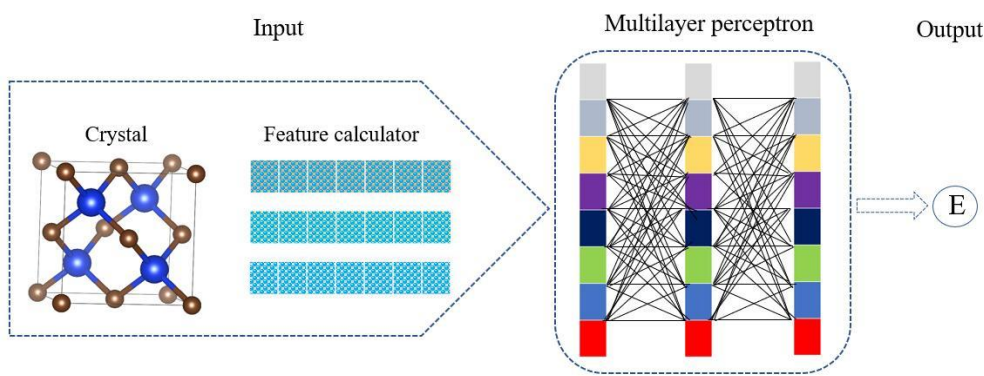

**Figure 3.** The graph representation of training network. Input is the feature vectors, output is the atomic energy. The color in figure is achieving a visual effect and it is insignificance.

The input to the network is a structure feature vector, which describes the local atomic environment of the crystal structure, and its dimensionality is determined by the structure features. By default, both hidden layers have the same number of nodes, and the output of the network corresponds to atomic energy. The choice of activation function, batch size, and optimization method for the neural network has been systematically studied and optimized, as detailed in our previous work[36]. CrySPAI provides recommended network parameters, though users can modify these parameters as needed.

The default parameters for the DNN module are pre-set. The minimum number of training samples for the initial model training iteration is set to 2000, and an early stopping method is employed to prevent overfitting. The root mean square error (RMSE) (Equation 1) between the model's predictions and the DFT-calculated results is used to monitor the training process, with a default convergence threshold of 8 meV/atom. Once the model has converged, it is stored and no longer updated for that specific calculation, and the model iteration loop is complete. To enhance the accuracy of the predicted model, a swarm algorithm, as described in our previous work[33], is integrated into CrySPAI. This algorithm helps to refine the energy calculations of structures, allowing CrySPAI to more easily identify stable structures.

$$\text{RMSE} = \frac{1}{n}(\frac{1}{N}\sum_{N}(E_{pre} - E_{DFT})^2)^{1/2} \tag{1}$$

where N is the total number of calculated structures, n is the number of atom in the calculated cell.

The converged model is used in the local optimization process of the EOA module, which operates outside of the model iteration loop shown in Figure 1. It is important to note that the model used in the local optimization process during the EOA module is in iterations is a model that has converged at the end of each loop, rather than a globally converged model.

### 4.2. Adaptive Volume Adjustment Algorithm

As volume is a key physical parameter in constructing crystal structures, CrySPAI allows users to provide the volume value of the target structure. Alternatively, the volume can be calculated using the adaptive volume adjustment algorithm.

In this algorithm, the structural volume is calculated as the sum of the volumes of all atoms in the structure, as given by Equation 2. *A* is an adjustment factor, with a default value of 1.2. The ionic radii, *R*, can be obtained either from the periodic table or from the "RCORE" parameter in the VASP pseudopotential file. CrySPAI offers several methods for users to specify the volume, including direct input of atomic or structural volume, extraction of ionic or atomic core radii from the periodic table or the VASP pseudopotential file, and other options.

$$V = \sum_i N_i * (\frac{4}{3}\pi R^3) * A \qquad (2)$$

## 5. Conclusions

In this work, we developed CrySPAI, a crystal structure prediction software based on artificial intelligence. CrySPAI integrates EOA for structure search, DFT for energy calculations, and DNN for modeling the relationship between structure and energy. These modules operate both independently and collaboratively, enabling efficient and accurate predictions of crystal structures.

To ensure a robust model, we used a diverse training dataset that includes structures from various crystal systems, stoichiometries, and cell sizes. This approach ensures CrySPAI's versatility in predicting a wide range of materials. Additionally, CrySPAI employs human-computer collaboration to refine and validate structure predictions, further improving overall accuracy. Currently, CrySPAI is tailored specifically for the prediction of crystal structures of inorganic materials, given their chemical compositions. While the present implementation is optimized for inorganic materials due to their distinct structural and energetic characteristics, the employed methods could theoretically be extended to organic or hybrid systems.

Although CrySPAI has demonstrated significant success in inorganic crystal structure predictions, further development are required.

(1) Dependence on training data. CrySPAI's prediction accuracy is heavily dependent on the quality and diversity of the training data used for DNN. To address this, we plan to expand the training dataset to cover a broader range of materials, , ensuring better generalization and improved performance.

(2) Scalability to complex systems. While CrySPAI is currently effective for inorganic materials, its application to highly complex systems, such as amorphous materials, organic materials, or systems with strong electron correlation effects, may require additional computational resources or tailored methods. Future iterations of CrySPAI will involve specific adaptations to handle these complex systems effectively.

(3) Generalizability to experimental conditions. CrySPAI predictions now are conducted under idealized computational conditions (e.g., 0K and no pressure). These may differ from experimental conditions, some phase transitions and defects states will also been thrown away. In the future versions, we will extend CrySPAI's capabilities to simulate dynamic processes, offering deeper insights into material behaviors under realistic conditions.

With these planed advancements, CrySPAI holds the potential to accelerate the discovery of new materials with tailored properties for a wide range of applications.

**Author Contributions:** All authors contributed to the study conception and design. Z.G. Wang and Y.G. Wang: methodology, software, validation, writing—original draft preparation, writing—review and editing and funding acquisition; Z.Y. Chen and Y. Yuan: software, writing—review and editing. All authors have read and agreed to the published version of the manuscript.

**Funding:** This research was supported by The Strategic Priority Research Program of Chinese Academy of Sciences (Grant No. XDB0500202), National Natural Science Foundation of China (Grant No.51802312), Key Research Program of Frontier Sciences, CAS (Grant NO.ZDBS-LY-7025), and Youth Innovation Promotion Association CAS.

**Institutional Review Board Statement:** Not applicable.

**Informed Consent Statement:** Not applicable.

**Data Availability Statement:** The data used to support the findings of this study are included within the article.

**Acknowledgments:** We thank our partners who provided all the help during the research process and the team for their great support.

**Conflicts of Interest:** The authors declare no conflicts of interest.

# References

1. Bergerhoff G.; Hundt, R.; Sievers R.; Brown I. D. The Inorganic Crystal Structure Data Base. *Journal of Chemical Information and Computer Sciences* **1983**, 23, 66-69.
2. VILLARS Pierre; CENZUAL Karin; GLADYSHEVSKII Roman; IWATA Shuichi. PAULING FILE - towards a holistic view. *Chemistry of Metals and Alloys* **2018**, 11, 43-76.
3. Curtarolo Stefano; Hart Gus L.W.; Nardelli Marco Buongiorno; Mingo Natalio; Sanvito Stefano; Levy Ohad. The high-throughput highway to computational materials design. *Nature Materials* **2013**, 12, 191-201.
4. Curtarolo Stefano; Morgan Dane; Ceder Gerbrand. Accuracy of ab initio methods in predicting the crystal structures of metals: A review of 80 binary alloys. *Calphad: Computer Coupling of Phase Diagrams and Thermochemistry* **2005**, 29, 163-211.
5. Hart Gus L.W.; Curtarolo Stefano; Massalski Thaddeus B.; Levy, Ohad. Comprehensive search for new phases and compounds in binary alloy systems based on platinum-group metals, using a computational first-principles approach. *Physical Review X* **2014**, 3, 1-33.
6. Doll, K.; Schön, J. C.; Jansen, M. Global exploration of the energy landscape of solids on the ab initio level. *Physical Chemistry Chemical Physics* **2007**, 9, 6128-6133.
7. Artem R Oganov; Colin W Glass. Evolutionary crystal structure prediction as a tool in materials design. *Journal of Physics: Condensed Matter* **2008**, 20, 064210.
8. Lyakhov Andriy O.; Oganov Artem R.; Stokes Harold T.; Zhu Qiang. New developments in evolutionary structure prediction algorithm USPEX. *Computer Physics Communications* **2013**, 184, 1172-1182.
9. Ji Min; Umemoto Koichiro; Wang Cai-Zhuang; Ho Kai-Ming; Wentzcovitch Renata M. Ultrahigh-pressure phases of $H_2O$ ice predicted using an adaptive genetic algorithm. *Physical Review B* **2011**, 84, 220105.

10. Wu Shunqing; Umemoto Koichiro; Ji Min; Wang Cai Zhuang; Ho Kai Ming; Wentzcovitch Renata M. Identification of post-pyrite phase transitions in $SiO_2$ by a genetic algorithm. *Physical Review B-Condensed Matter and Materials Physics* **2011**, 83, 6-9.

11. S Q Wu; M Ji; C Z Wang; M C Nguyen; X Zhao; K Umemoto; R M Wentzcovitch; K M Ho. An adaptive genetic algorithm for crystal structure prediction. *Journal of Physics: Condensed Matter* **2013**, 26, 035402.

12. Wang Yanchao; Lv Jian; Zhu Li; Ma Yanming. Crystal structure prediction via particle-swarm optimization. *Physical Review B-Condensed Matter and Materials Physics* **2010**, 82, 1-8.

13. Wang Yanchao; Lv Jian; Zhu Li; Ma Yanming. CALYPSO: A method for crystal structure prediction. *Computer Physics Communications* **2012**, 183, 2063-2070.

14. Pickard Chris J.; Needs R. J. Ab initio random structure searching. *Journal of Physics: Condensed Matter* **2011**, 23, 053201.

15. Lonie David C.; Zurek Eva. XtalOpt: An open-source evolutionary algorithm for crystal structure prediction. *Computer Physics Communications* **2011**, 182, 372-387.

16. Ouyang Runhai. Exploiting Ionic Radii for Rational Design of Halide Perovskites. *Chemistry of Materials* **2020**, 32, 595-604.

17. Christopher J. Bartel; Christopher Sutton; Bryan R. Goldsmith; Runhai Ouyang; Charles B. Musgrave; Luca M. Ghiringhelli; Matthias Scheffler. *Science Advances* **2019**, 5, eaav0693.

18. Kusne, A. Gilad; Yu, Heshan; Wu Changming; Zhang, Huairuo; Hattrick-Simpers Jason; DeCost Brian; Sarker Suchismita; Oses Corey; Toher Cormac; Curtarolo Stefano; Davydov Albert V.; Agarwal Ritesh; Bendersky Leonid A.; Li Mo; Mehta Apurva; Takeuchi Ichiro. On-the-fly closed-loop materials discovery via Bayesian active learning. *Nature Communications* **2020**, 11, 5966.

19. Gabriel R Schleder; Antonio C M Padilha; Carlos Mera Acosta; Marcio Costa; Adalberto Fazzio. From DFT to machine learning: Recent approaches to materials science-A review. *JPhys Materials* **2019**, 2, 032001.

20. Schmidt Jonathan; Marques Mário R.G.; Botti Silvana; Marques Miguel A.L. Recent advances and applications of machine learning in solid-state materials science. *npj Computational Materials* **2019**, 5, 83.

21. Wei Jing; Chu Xuan; Sun Xiang-Yu; Xu Kun; Deng Hui-Xiong; Chen Jigen; Wei Zhongming; Lei Ming. Machine learning in materials science. *InfoMat* **2019**, 1, 338-358.

22. Bohayra Mortazavi; Evgeny V Podryabinkin; Ivan S Novikov; Stephan Roche; Timon Rabczuk; Xiaoying Zhuang; Alexander V Shapeev. Efficient machine-learning based interatomic potentialsfor exploring thermal conductivity in two-dimensional materials. *JPhys Materials* **2020**, 3, 02LT02.

23. Vasudevan, Rama and Pilania, Ghanshyam and Balachandran, Prasanna V. Machine learning for materials design and discovery. *Journal of Applied Physics* **2021**, 129, 070401.

24. Tian Xie; Jeffrey C. Grossman. Crystal Graph Convolutional Neural Networks for an Accurate and Interpretable Prediction of Material Properties. *Physical Review Letters* **2018**, 120, 145301.

25. Chi Chen; Weike Ye; Yunxing Zuo; Chen Zheng; Shyue Ping Ong. Graph Networks as a Universal Machine Learning Framework for Molecules and Crystals. *Chemistry of Materials* **2019**, 31, 3564-3572.

26. Yang Yuan; Ziyi Chen; Tianyu Feng; Fei Xiong; Jue Wang; Yangang Wang; Zongguo Wang. Tripartite interaction representation algorithm for crystal graph neural networks. *Scientific Reports* **2024**, 14, 24881.

27. Chi Chen; Shyue Ping Ong. A universal graph deep learning interatomic potential for the periodic table. *Nature Computational Science* **2022**, 2, 718-728.

28. Bowen Deng; Peichen Zhong; KyuJung Jun; Janosh Riebesell; Kevin Han; Christopher J. Bartel; Gerbrand Ceder. CHGNet as a pretrained universal neural network potential for charge-informed atomic modelling. *Nature Machine Intelligence* **2023**, 5, 1031-1041.

29. Fankai Xie; Tenglong Lu; Sheng Meng; Miao Liu. GPTFF: A high-accuracy out-of-the-box universal AI force field for arbitrary inorganic materials. *Science Bulletin* **2024**, 69, 3525-3532.

30.  Artrith Nongnuch; Urban Alexander. An implementation of artificial neural-network potentials for atomistic materials simulations: Performance for $TiO_2$. *Computational Materials Science* **2016**, 114, 135-150.

31.  Zhang Linfeng; Lin De-Ye; Wang Han; Car Roberto; E Weinan. Active learning of uniformly accurate interatomic potentials for materials simulation. *Physical Review Materials* **2019**, 3, 023804.

32.  Zhang, Linfeng; Han, Jiequn; Wang, Han; Car, Roberto; E, Weinan. Deep Potential Molecular Dynamics: A Scalable Model with the Accuracy of Quantum Mechanics. *Physical Review Letters* **2018**, 120, 143001.

33.  Wang Junjie; Gao Hao; Han Yu; Ding Chi; Pan Shuning; Wang Yong; Jia Qiuhan; Wang Hui-Tian; Xing Dingyu; Sun Jian. MAGUS: machine learning and graph theory assisted universal structure searcher. *National Science Review* **2023**, 10, nwad128.

34.  Cheng Guanjian; Gong Xin Gao; Yin Wan Jian. Crystal structure prediction by combining graph network and optimization algorithm. *Nature Communications* **2022**, 13, 1492.

35.  Merchant Amil; Batzner Simon; Schoenholz Samuel S.; Aykol Muratahan; Cheon Gowoon; Cubuk Ekin Dogus. Scaling deep learning for materials discovery. *Nature* **2023**, 624, 80-85.

36.  Zhi-wei Liu; Zong-guo Wang; Jia-long Guo; Yan-gang Wang. Deep Learning Method for Crystal Structure Prediction. *Computer Systems & Applications* **2021**, 30, 40-49.

37.  Zhao Xin; Shu Qiang; Nguyen Manh Cuong; Wang Yangang; Ji Min; Xiang Hongjun; Ho Kai-Ming; Gong Xingao; Wang Cai-Zhuang. Interface Structure Prediction from First-Principles. *The Journal of Physical Chemistry C* **2014**, 118, 9524-9530.

38.  Perdew John P.; Burke Kieron; Ernzerhof Matthias. Generalized gradient approximation made simple. *Physical Review Letters* **1996**, 77, 3865-3868.

39.  Artrith Nongnuch; Urban, Alexander; Ceder Gerbrand. Efficient and accurate machine-learning interpolation of atomic energies in compositions with many species. *Physical Review B* **2017**, 96, 014112.

40.  Zhiwei Liu; Jialong Guo; Ziyi Chen; Zongguo Wang; Zhenan Sun; Xianwei Li; Yangang Wang. Swarm intelligence for new materials. *Computational Materials Science* **2022**, 214, 111699.